\documentclass[aps,prb,amsmath,amssymb,reprint]{revtex4-1}

\usepackage{graphicx}
\renewcommand{\deg}{$^{\circ}$ }

\begin{document}

\title{Dynamic dependence to domain wall propagation through artificial spin ice}

\author{D.M. Burn}
\author{M. Chadha}
\author{W.R. Branford}
\affiliation{Department of Physics, Imperial College London, London SW7 2BZ, United Kingdom}
\date{\today}

\begin{abstract}
Domain wall propagation dynamics have been studied in nanostructured artificial kagome spin ice structures. A stripline circuit has been used to provide localised pulsed magnetic fields within the artificial spin ice structure. This provides control of the system through electrically assisted domain wall nucleation events. Synchronisation of the pulsed fields with additional global magnetic fields and the use of a focussed magneto-optical Kerr effect magnetometer allows our experiments to probe the domain wall transit through an extended ASI structure. We find that the propagation distance depends on the driving field revealing field driven properties of domain walls below their intrinsic nucleation field.
\end{abstract}

\pacs{}
\keywords{}
\maketitle



Magnetic meta-materials such as artificial spin ice show behaviour arising from complex geometrical structuring in addition to the original material properties.\cite{RMP_85_1473, J.Phys.Cond.Matt_25_363201} In these systems it is the combination of magnetic charge interactions and topological constraints determine the magnetisation behaviour of the system. Artificial spin ice structures consisting of arrays of magnetic nanobars provide a 2D analogue to explore frustrated magnetic phenomena.\cite{Nat_439_303, N.Phys_7_68} These systems are of fundamental scientific interest\cite{N.Phys_6_359, Phil.Soc.R.Trans_370_5767, PRL_111_057204, Science_335_1597, NJP_15_035026, PRB_77_094418, Nat_500_553, APL_101_112404} and have even been identified as potentials for novel neural network or processing technologies.\cite{Nat_439_303, NJP_13_023023, Nat.Phys_6_786}

Magnetisation reversal in artificial spin ice structures composed of interconnected magnetic bars can be described by an ensemble of magnetic domain wall (DW) processes. The creation, annihilation and propagation of these DWs throughout the system leads to magnetisation reversal within the bars as well as the transport of both magnetic and topological charges throughout the system. The conservation of both magnetic and topological charge provides constraints on the creation and annihilation of DWs in the system. This reveals the physical significance of the finer details of the micromagnetic DW structure such as its chirality or topological makeup when the DW interacts with a complex magnetic structure.\cite{PRL_107_167201, PRL_95_197204, N.Phys_9_505, NJP_17_013054, Science_335_1597, NJP_15_035026, PRL_105_187206} 

The majority of our understanding of the magnetisation behaviour in artificial spin ice systems is based on experiments combining thermal and quasi-static magnetic fields applied to the entire system.\cite{Sci.Rep_4_5702, APL_101_112404, PRL_109_037203, PRL_105_047205, PRB_77_094418}  The role of DWs have been typically investigated based on their natural occurrence,\cite{NJP_14_045010} when an applied field exceeds the nucleation field which is typically lower at the edges of the structures. This approach is therefore limited in that we can only investigate the internal behaviour of the system once a process related to the edge of the system takes place. 

In this investigation the localised injection of DWs along the length of a lithographically patterned microstrip is employed.\cite{PRL_96_197207, JAP_93_8430} Here the pulsed field DW injection technique allows control over the DW nucleation location within the system leading to significant experimental advantages as the DW nucleation process can be separated from a global applied field. Firstly, this allows the behaviour of DWs in the system to be investigated in a wider field range, even at lower fields than their nucleation field. Secondly, this allows the magnetisation dynamics in the system to be explored. This is of great interest in the artificial spin ice system and understanding of the behaviour in a dynamic context is necessary for any future technological applications.

\begin{figure}[b]
	\centering
	\includegraphics[width=8.5cm]{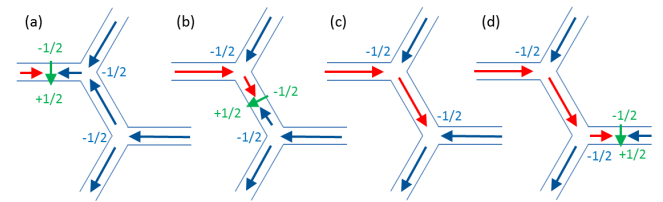}
	\caption{Simple model of DW progation and annihilation based on topological constraints in multiple vertices.}
	\label{cartoon_topology}
\end{figure}

Our understanding of the propagation path of a DW through a series of vertex structures can be explained through topological considerations. Figure \ref{cartoon_topology} shows a two-vertex section of an artificial spin ice structure with arrows representing the magnetisation orientation in each bar and with topological defects pinned to the edges of the structure associated with both the DWs and the vertices. 

Figure \ref{cartoon_topology} shows the evolution of a down-chirality DW incident upon a vertex with initial magnetisation saturated to the left. During the interaction the $-1/2$ topological defect initially belonging to the DW becomes pinned on the upper edge of the vertex. The $+1/2$ from the DW follows the edge of the structuring and pairs with the $-1/2$ initially associated with the vertex. This new defect pair corresponds to a DW which is able propagate along the lower branch at the vertex. The similar +ve magnetic charges of the initial DW and the vertex provide a repulsive force which means there is an energy barrier associated with this process which can be overcome through the application of an applied magnetic field.

In figure \ref{cartoon_topology}(b) the -ve charge of the second vertex now provides an attractive force on the positively charged DW. Here the topological charges on the lower side of the nanobar are opposite and therefore unwind when they meet. The DW annihilates resulting in the two-in one-out state illustrated in figure \ref{cartoon_topology}(c) where just a $-1/2$ topological defect from the incident DW now remains pinned at the vertex. In this model, the DW no longer exists in the system.

Figure \ref{cartoon_topology}(d) shows how a new DW can be injected into the lower horizontal nanobar based on the state shown in figure \ref{cartoon_topology}(c). The lower edge of the vertex contains zero topological charge which is separated to form two defects of $+1/2$ and $-1/2$ respectively. The $+1/2$ forms a pair with the pre-existing $-1/2$ defect on the upper edge of the vertex and represents a DW which can propagate along the nanobar whilst the remaining $-1/2$ defect remains at the lower edge of the vertex. This process involves the separation of two opposite magnetic charges which gives rise to an energy barrier which needs to be overcome to complete this process.

The series of interactions illustrated in figure \ref{cartoon_topology} shows how an initial DW with down-chirality can interact with two verticies in an artificial spin ice structure resulting in a down-chirality DW in a subsequent nanobar with similar geometry. The repeat of this process is consistent with reversal events following from one another. A similar series of interactions would also take place for up-chirality DW which would propagate along the upper branch at the first vertex. This would be followed by an equivalent annihilation event at the second vertex and the availability to nucleate a further up-chirality DW in the final horizontal nanobar. In all cases, a sizable magnetic field is required to supply the energy to overcome the energy barriers associated with moving like charges towards one another, and separating zero charge into a positive and negative charge pair.

This current understanding is based on the system which maintains a minimum energy micromagnetic spin configuration. In this quasi-static regime energy must be supplied to overcome energy barriers associated with the transitions which nucleate the DW in new nanobars throughout the system. In this work we consider the higher energy processes involved with dynamic propagating DWs and show deviations from our understanding of these processes in the quasi-static regime.

By varying the bias field that drives DW motion we investigate the importance of DW dynamics during propagation through an artificial spin ice structure with fields applied along the armchair geometry. Combining localised DW injection with a MOKE measurements with a localised magnetisation probe we also infer on the lengthscales of propagation through the system at these fields. Critically, our experiments probe the DW propagation behaviour in fields below the intrinsic DW nucleation field for these structures.



\begin{figure}[bp]
	\centering
	\includegraphics[width=8.5cm]{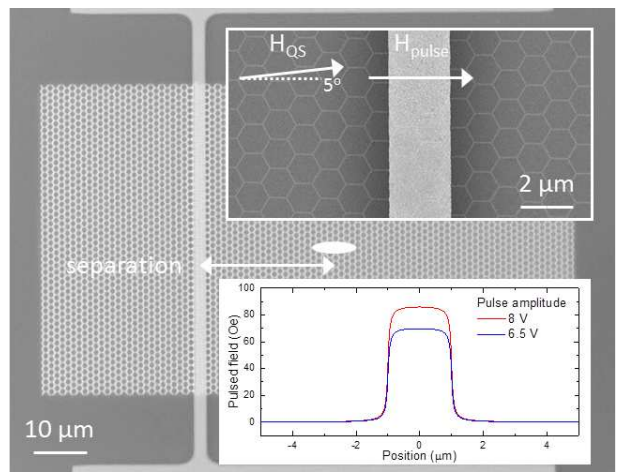}
	\caption{Image of a interconnected array of nanowires with a Au stripline for applying localised pulsed magnetic fields. 700~nm x 150 nm}
	\label{image}
\end{figure}

Artificial spin ice structures consisting of interconnected NiFe nanowires were fabricated in a kagome geometry using electron-beam lithography and thermal evaporation. The bars were 700~nm x 150~nm in dimension and were 10~nm thick. Further details about the patterning of the magnetic structures can be found elsewhere \cite{PRB_92_214425}. A 2~$\mu$m wide Cr(5~nm)/Au(50~nm) microstrip was added in a second lithographic process and is shown in figure \ref{image} along with the simulated field profile expected from the microstrip.\cite{JAP_85_7849}

The magnetisation reversal in the system was investigated using magneto-optical Kerr effect (MOKE) magnetometry in the longitudinal geometry. Here, a focussed laser spot with a  $\sim10$~$\mu$m elongated footprint provided a localised probe of the magnetisation reversal as illustrated in figure \ref{image}. The combination of quasi-static and pulsed magnetic fields, supplied from external coils and from the microstrip respectively, were used to drive the magnetisation reversal in the sample. The Kerr signal was averaged over 50 field cycles in each measurement. 

By introducing a 5\deg angular offset between ASI structuring and the applied field direction the magnetisation reversal associated with DW nucleation events and DW propagation through the system can be distinguished by the reversal field.\cite{PRB_92_214425} Additionally, by varying the contribution from the quasi-static and pulsed fields, behaviour from quasi-static energy dependent magnetisation reversal processes and time-dependent magnetisation processes have been investigated.\cite{PRB_88_104422, PRL_96_197207}



\begin{figure}[bp]
	\centering
	\includegraphics[width=8.5cm]{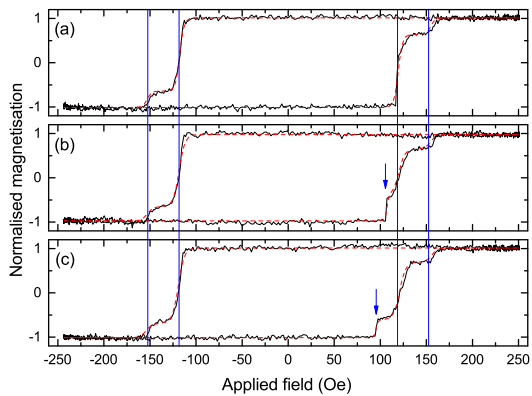}
	\caption{MOKE hysteresis loops showing the magnetisation reversal driven by (a) a quasi-static magnetic field and (b) and (c) with the addition of a pulsed magnetic field with triggering indicated by the arrows. The loops show the behaviour 9V 20ns pulses with the laser at the stripline. The fitted lines show a model fit to the data.}
	\label{mh}
\end{figure}

Initially the magnetisation behaviour was investigated with a 1~Hz sinusoidal quasi-static applied magnetic field and is shown in figure \ref{mh}(a). Two transitions in the magnetisation occur at two distinct and relatively sharp reversal fields despite averaging over 50 field cycles and over multiple nanowires within the illuminated laser footprint. The two steps indicate the combination of two reversal processes occurring during the magnetisation reversal associated with the nucleation field of a DW from a vertex and the field required for a pre-existing DW to propagate through a vertex.\cite{PRB_92_214425}

By introducing additional pulsed magnetic fields, figures \ref{mh}(b) and (c) show modified hysteresis loops where the arrows indicate the triggering of the pulsed field within the quasi-static field cycle resulting in an increase in the magnetisation. At this point, the combination of pulsed and quasi-static fields locally overcome the reversal field leading to the injection of magnetic DWs at the stripline. The additional pulsed field induced reversal results in a reduced magnetisation reversal at the lower quasi-static field and the higher field quasi-static reversal remains unchanged.

The magnetisation reversal combines multiple reversal processes which can be modelled as a summation of $tanh$ functions.\cite{IEEE.Trans.Magn_36_404} The lines in figure \ref{mh} show a model fit to the data where each transition is parameterised by a reversal field, the relative change in magnetisation and a parameter representing the shape of that transition. The quasi-static reversal fields are symmetric with increasing and decreasing field and share fitting parameters.

The shape of the hysteresis loop representing the increase in magnetisation at the lower reversal field also shows some broadening when a pulsed field is present. This could represent a modified reversal field due to a partially-reversed magnetisation state following the pulse. This will be discussed in more detail later.

\begin{figure}[bp]
	\centering
	\includegraphics[width=8.5cm]{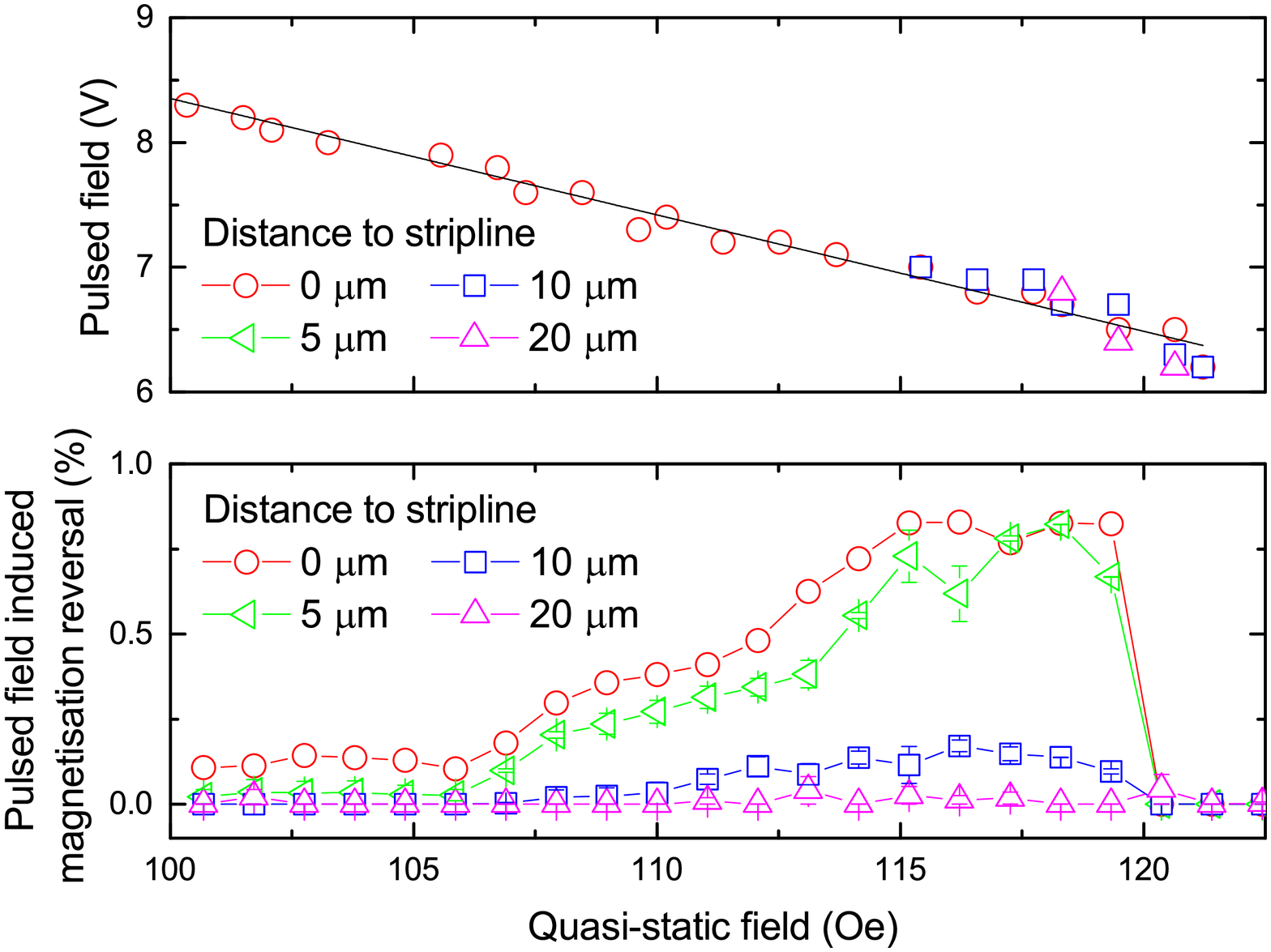}
	\caption{a) Minimum pulsed voltage required to lead to magnetisation reversal when pulses are triggered at different QS Bias fields. b)Pulsed field induced magnetisation reversal as a function of quasi-static bias field measured at various separations from the DW injection stripline. Pulses are 20~ns and 7.5~V}
	\label{stripline_calibration}
\end{figure}

The pulsed-field induced magnetisation reversal depends on both the pulsed field voltage and the triggering point within the quasi-static field cycle. Figure \ref{stripline_calibration}(a) shows the minimum pulse voltage required to result in the additional pulsed-field induced magnetisation reversal steps in figure \ref{mh} (b and c). This is plotted as a function of the quasi-static field at the point of pulsed field triggering which can be considered as a static bias field on the timescales of the pulsed field. The linear decrease represents the contributions to the total field at the stripline where an increase in pulsed field amplitude from a greater pulsed voltage allows magnetisation reversal to take place at a lower quasi-static field. A linear fit to this data provides a calibration of the stripline which produces $10.8 \pm 0.3$~Oe/V. For quasi-static fields greater than 122~Oe the magnetisation reversal is driven purely by the quasi-static field. Therefore the effect of the pulsed field in this field regime cannot be distinguished.

Figure \ref{stripline_calibration}(a) also compares the difference in behaviour when the laser spot is positioned 0, 10 and 20~$\mu$m away from the stripline. All the points fall on the same line indicating the reversal process at the stripline is not affected by the measurement position. However, when measurements are performed with greater separation, magnetisation reversal is only observed when the pulsed field triggering occurs at larger quasi-static fields. This feature allows us to probe the motion of the DWs through an extended region of the ASI system.


For all measurement positions the combination of pulsed and quasi-static fields still result in magnetisation reversal at the stripline. This reversal is due to the locaslied injection of magnetic DWs at the stripline which propagate along the nanobars reversing the magnetisation near the stripline. At greater distances from the stripline only the quasi-static field drives the DW propagation and the magnetisation reversal represents the dynamic behaviour of DWs in fields below their nucleation field. 


Pulse lengths of 150~ns and 20~ns had little influence on the DW injection process so the quasi-static bias field dependence on DW propagation has been further investigated with 20~ns long, 7.5~V pulses which are sufficient to inject DWs when biased with a field greater than 108~Oe. Figure \ref{stripline_calibration}(b) shows the pulsed-field-induced magnetisation reversal contribution as a function of the quasi-static bias field for various measurement positions. This is found from the ratio in magnetisation change from the pulsed and quasi-static fields in the hysteresis loops.

Measurements at the stripline location show a large pulsed-field induced magnetisation change which is most significant when a large quasi-static field is used to drive the DW propagation. Here, the result represents DWs which reverse the magnetisation in nanobars near where they are nucleated.

When the quasi-static bias field is reduced below 115~Oe the magnetisation change decreases. This indicates that the proportion of pulsed field induced reversal events within the region probed by the laser spot is reduced. This can be explained by DWs which are not able to propagate so far through the structure in the lower fields. 


Measurements at greater distances from the stripline show magnetisation reversal driven purely by the quasi-static field (see field profiles in figure \ref{image}). Here a lower magnetisation change is found as injected DWs must propagate through a greater number of nanobars and vertices before reaching the probed region. This means that DW pinning is more likely, but at high quasi-static bias fields, DW propagation up to 15~$\mu$m is still observed.

\begin{figure}[tbp]
	\centering
	\includegraphics[width=8.5cm]{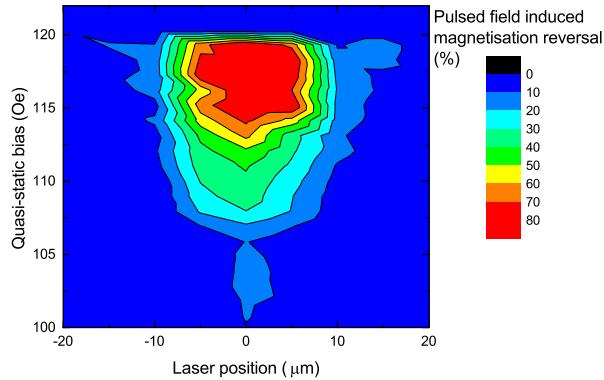}
	\caption{Pulsed field induced magnetisation reversal measured as a function of quasi-static bias field and measurement position from the stripline. 20~ns Pulses at 7.5~V }
	\label{pulsed_contour}
\end{figure}

The DW propagation distance through the structure is more clearly shown in in figure \ref{pulsed_contour} where the pulsed-field-induced magnetisation reversal is plotted as a function of the measurement position and the quasi-static field. Here a strong reversal is centered around the position of the stripline at 0~$\mu$m which becomes more significant with greater quasi-static fields. Again, as the fields approach 120~Oe the pulsed-field-induced reversal becomes indistinguishable from the quasi-static field driven reversal.

With the large quasi-static fields, the distance over which the reversal can be detected is much greater than for the smaller quasi-static fields resulting in a triangular shape in figure \ref{pulsed_contour}. This shows how the propagation of a DW through the artificial spin ice structure depends on the driving field applied to the DW. With low fields the propagation distance is limited as multiple nanobars and interconnecting verticies are encountered and provide pinning sites. However, with greater fields the pinning from these become less significant allowing the DW propagation over a greater number of nanobars and vertices.

The simplified model of the DW path illustrated in figure \ref{cartoon_topology} relies on the external driving field exceeding the nucleation field for a DW in these structures. However, our results reveal that DWs are able to propagate below this field with a driving field dependence to their propagation length through the system.

The propagation at a reduced field can be explained by considering the energetics associated with a propagating DW. When considering the annihilation process between the DW and vertex in figure \ref{cartoon_topology}(b) and (c) the energy associated with the DW can be used to assist with the nucleation of the DW in figure \ref{cartoon_topology}(d). This additional energy would mean less is required externally from the driving field.

The propagation length dependence can also be explained in terms of the energetics of the dynamically propagating DWs. With a greater quasi-static driving field the DWs travel with a greater energy. This means that they can encounter, and overcome a greater number energy barriers associated with the verticies before becoming pinned. The lower energy DWs become pinned after fewer verticies and therefore travel a reduced distance through the artificial spin ice structuring.



In conclusion, we have probed the magnetisation reversal in artificial spin ice systems through focussed MOKE magnetometry. The combination of pulsed and quasi-static magnetic fields allow for the injection of DWs and the study of their dynamic interactions with the geometrical structuring for a range of DW driving fields. Our results demonstrate control over the location of injected DWs within artificial spin ice structures and how DW propagation distance depends on the external driving field.

Existing quasi-static models based on the manipulation of magnetic and topological charges throughout the system do not predict a length dependence to the propagation. Our results, probing DWs in the dynamic regime with a range of driving fields suggest changes for DWs arriving at vertices in a higher energy state.



\end{document}